\documentclass[paper=a4, 12pt]{article}	

\usepackage[english]{babel}															
\usepackage[protrusion=true,expansion=true]{microtype}				
\usepackage{amsmath,amsfonts,amsthm, amssymb,amsbsy}										
\usepackage[pdftex]{graphicx}														
\usepackage{epstopdf}																	
\usepackage{subfig}																		
\usepackage{booktabs}																	
\usepackage{latexsym}
\usepackage{natbib}

\usepackage{authblk}
\usepackage{amsmath}
\usepackage{amsmath,amsfonts,amssymb,amsthm,epsfig,epstopdf,url,array}
\theoremstyle{plain}

\theoremstyle{definition}

\usepackage{verbatim}
\usepackage{fancyvrb}

\newcommand{\diag}{\mbox{diag}}
\newcommand{\Var}{\mbox{Var}}
\newcommand{\Cov}{\mbox{Cov}}
\newcommand{\cov}{\mbox{Cov}}
\newcommand{\E}{\mbox{E}}

\newcommand{\ve}[1]{\mbox{\boldmath ${#1}$}}
\newcommand{\vesub}[2]{\mbox{{\boldmath ${#1}$}$_{#2}$}}
\newcommand{\vesup}[2]{\mbox{{\boldmath ${#1}$}$^{#2}$}}
\newcommand{\vess}[3]{\mbox{{\boldmath ${#1}$}$_{#2}^{#3}$}}
\newcommand{\hve}[1]{\hat{\ve{#1}}}
\newcommand{\hvesub}[2]{\hat{\ve{#1}}_{#2}}

\newcommand{\bs}{\boldsymbol}

\title{Regression Analysis for Multivariate Dependent Count Data Using Convolved Gaussian Processes
}
  \author[1]{A'yunin Sofro}
\author[2]{Jian Qing Shi \thanks{Corresponding author: j.q.shi@ncl.ac.uk}}	
\author[3]{Chunzheng Cao}
\smallskip
\affil[1]{Department of Mathematics, Surabaya State University, Indonesia }
\affil[2]{School of Mathematics, Statistics and Physics, Newcastle University, UK}
\affil[3]{School of Mathematics \& Statistics, Nanjing University of Information Science  and Technology, China}
\date{\today}

\begin{document}

\maketitle


\begin{abstract}
Research on Poisson regression analysis for dependent data has been developed rapidly in the last decade. One of difficult problems in a multivariate case is how to construct  a cross-correlation structure and at the meantime make sure that the covariance matrix is positive definite.  To address the issue, we propose to use convolved Gaussian process (CGP) in this paper. The approach provides a semi-parametric model  and offers a natural framework for  modeling common mean structure and covariance structure simultaneously. The CGP enables the model to define different covariance structure for  each component of the response variables. This flexibility ensures the model to cope with data coming from different resources or having different data structures, and thus to provide accurate estimation and prediction.  In addition, the model is able to accommodate  large-dimensional covariates.
The definition of the model, the inference and the implementation, as well as its asymptotic properties, are discussed. Comprehensive numerical examples with both simulation studies and real data are presented.

\end{abstract}

\section{Introduction}
Regression analysis for dependent non-Gaussian data has been developed rapidly in the last several decades. We will focus on depedent count data in this paper. One way is to extend the conventional   Poisson regression model by considering a covariance stucture. However, the problem of modelling  becomes more complex when there is more than one response variable. We illustrate the challenges using the example of dengue fever and malaria data that we will discuss in details later in this paper. The outputs are the number of cases of  dengue fever and malaria occurred in different regions in East Java in Indonesia.
Both diseases are transmitted by a virus via mosquitoes and occur often in tropical regions particularly in developing countries. They have similar signs and symptoms. The outbreak of the diseases depends on many factors such as living condition and healthy behaviour. The data is spatially correlated due to the movement of population, analogues of the environment and the healthy behaviour, etc. The study for such problems focuses on the following three aspects. First of all, we want to study how the count of cases depends on a set of covariates. A parametric model is usually used since it can provide a physical explanation on the relationship between the disease and the covariates. Secondly, we are interested in finding the structure of spatial correlation of the depend data for each disease and further to find the geographical patterns. This provides a tool in epidemic study. Due to the nature of the problem, it requires a flexible covariance model and ideally the covariance structure and the pattern can be learned from data rather than an assumption given in advance. Thirdly, we want to study similar diseases or response variables at the same time. We are interested in knowing if there are similar geographical patterns for those diseases and how they are spatially correlated and cross-correlated. The findings will provide important information for policy making on how to control the spread and transmission of the  diseases.

Poisson regression  analysis for an univariate count response variable with correlation structure has been studied by many researchers. The intrinsic conditional autoregressive (ICAR) model is one of the popular methods which was introduced by \citet{Besag95}. This method has been extended into a spatial or temporal correlated generalized linear mixed model \citep{Sun00,MacNab01,Martinez-Beneito08,Silva08}. A generalized linear mixed model using prior distribution for spatially structured random effect is an alternative way, see \citet{
Banerjee04}.  \citet{Rue05} and \citet{Mohebbi11} demonstrated how to apply the methods to analyse cancer data. However, based on extensive studies by \citet{Wall04}, the spatially correlated structure of ICAR approach is too complicated, involving complex implementation and lack of physical explanation. \citet{Martinez-Beneito13} has also pointed out that preliminary knowledge and a good understanding are needed in determining and investigating the effect of the choice of precision for the covariance matrix.  Thus, it is essential to develop a more flexible method to model the spatial correlation. One alternative is to use a Gaussian process (GP) prior (or kriging under spatial statistics, see \citet{Diggle98}) to model the covariance structure (see e.g. \citet{Rasmussen06} and \citet{Shi11}). This is a nonparametric approach, providing a flexible method on modeling covariance structure. The Bayesian framework with GP priors with different covariance functions provides flexibility on fitting data with different degrees of nonlinearity and smoothness. It can also cope with multi-dimensional covariates. Some recent development can be found in e.g. \citet{Gramacy12} and \citet{Wang14}.

For the problem involved multivariate response variables, we need to model covariance structure for each component as well as cross-covariance between them. The challenge here is how to find a model which can model the covariance and cross-variance flexibly, subject to the condition that the overall covariance function is positive definite. Several methods have been proposed, for example,   two-fold CAR model \citep{Kim01} and  multivariate CAR (MCAR) \citep{Gelfand03}.  \cite{Jin05} proposed a general framework for MCAR by using a conditional approach
$p(\tau_{1},\tau_{2})=p(\tau_{1}\mid\tau_{2})p(\tau_{2})$, where $\tau_1$ and $\tau_2$ stand for the two components. As we pointed out before, the CAR model is useful for some problems but is less efficient for a general use. \citet{Crainiceanu12} also used the idea of conditional distribution but the covariance structure is modeled by a GP prior. It provides a promising result for some types of problem. However the covariance structure of  $\tau_1$ depends on the covariance structure of $\tau_2$. If those two components have very different covariance structures, the model may be failed. The performance also depends on the ordering of the components. An additional problem is that it is not easy to extend it to cases with more than two components.

In this paper we propose to use convolved GP (CGP) \citep{Boyle05} and provides a general framework on modeling individual covariance structure for each component and, at the same time, modeling cross-covariance for multivariate count data. The method can be easily extended to deal with multivariate case with any dimension. It inherits nice properties of GP model, for example,  it offers a semiparametric regression model for Poisson data with multivariate responses;  it models mean structure and covariance structure simultaneously; and it enables us to handle a large dimensional covariates.

This paper is organized as follows. In section 2, we will discuss how to construct multivariate dependent Gaussian processes using convolution. We will then explain the details how to define a multivariate CGP for dependent count data. The details of inference including estimation, prediction and  asymptotic theory will also be provided in the section. Comprehensive  simulation studies and real data  applications  will be discussed in Section 3. The final conclusive remarks will be given in Section 4.

\section{\label{}Multivariate CGP model for Dependent Count Data}

\subsection{\label{} Multivariate Convolved Gaussian Processes}

We first introduce Multivariate Convolved Gaussian Processes (MCGP) and defer the definition of the main model to the next subsection.
Let $\gamma(\bs{x})$ be a Gaussian white noise $\gamma(\bs{x}) \overset{iid}{\sim} \mathcal{N}(0,\sigma^{2})$ and $h(\bs{x})$ be a smoothing kernel for $\bs{x} \in \mathcal{R}^{p}$. We can construct a CGP $\eta(\bs{x})$ \citep{Boyle05,Shi11} as
\begin{eqnarray}\label{41a}
\eta(\bs{x})&=&h(\bs{x})\star \gamma(\bs{x})\nonumber
=\int h(\bs{x}-\bs \alpha) \gamma(\bs \alpha) \textup{d} \bs{\alpha}  = \int h(\bs{\alpha}) \gamma(\bs{x}-\bs{\alpha}) \textup{d}\bs\alpha,
\end{eqnarray}
where `$\star$' denotes convolution. We denote it by
\begin{eqnarray}\label{41}
\eta(\bs{x}) \sim  \textup{CGP} (h(\bs{x}),\gamma(\bs{x})).
\end{eqnarray}
For example, if we choose a smooth kernel $h(\bs{x})$ as
$
h(\bs{x})= v \exp\left \{ -\frac{1}{2}(\bs{x}-\bs \mu)^{T}\bs{A} (\bs{x}-\bs \mu)\right \},
$
then the CGP $\eta(\bs{x})$ defined in \eqref{41a} is equivalent to  a GP with zero mean and  the following covariance function
\begin{eqnarray}\label{42}
k(\bs{x}_{i},\bs{x}_{j})&=&\pi^{p/2}v^{2} \lvert \bs{A} \rvert ^{-1/2}\exp \left \{-\frac{1}{4}(\bs{x}_{i}-\bs{x}_{j})^{T}\bs{A} (\bs{x}_{i}-\bs{x}_{j})  \right \},
\end{eqnarray}
for any $\vesub{x}{i}, \vesub{x}{j} \in  \mathcal{X} \subset \mathcal{R}^p$, where $v$ and $\bs{A}$ are parameters. This is the squared exponeential covariance function.

To define a bivariate CGP, we first define three independent Gaussian white noises, namely $\gamma_{0}(\bs{x}), \gamma_{1}(\bs{x})$ and $\gamma_{2}(\bs{x})$. Using them, we  construct four CGPs as follows:
\begin{equation}\label{e2}
\xi_{1}(\bs{x}) \sim  \textup{CGP} (h_{1}(\bs{x}),\gamma_{0}(\bs{x})), \quad
\xi_{2}(\bs{x}) \sim  \textup{CGP} (h_{2}(\bs{x}),\gamma_{0}(\bs{x}))
\end{equation}
and
\begin{equation}\label{e1}
\eta_{1}(\bs{x}) \sim  \textup{CGP} (g_{1}(\bs{x}),\gamma_{1}(\bs{x})), \quad
\eta_{2}(\bs{x}) \sim   \textup{CGP} (g_{2}(\bs{x}),\gamma_{2}(\bs{x})),
\end{equation}
where $g_a(\bs x)$ and $h_a(\bs x) \ (a=1, 2)$ are smoothing kernels.  It is clear that $\eta_{1}(\bs{x})$ and $\eta_{2}(\bs{x})$ are independent,  $\xi_{1}(\bs{x})$ and $\xi_{2}(\bs{x})$ are dependent but are independent from $\eta_{1}(\bs{x})$ and $\eta_{2}(\bs{x})$. Using  those four CGPs we can define bivariate dependent GPs as
\begin{equation}\label{43}
\tau_{a}(\bs{x}) = \xi_{a}(\bs{x})+\eta_{a}(\bs{x}),\quad a= 1,~2.
\end{equation}
Based on equation in \eqref{43}, the dependency between $\tau_{1}(\bs{x})$ and $\tau_{2}(\bs{x})$ is modeled by $\xi_{1}(\bs{x})$ and $\xi_{2}(\bs{x})$, while the individual characteristics are modeled by $\eta_{1}(\bs{x})$ and $\eta_{2}(\bs{x})$. Since the covariance structure can be modeled by different smoothing kernels $g_a(\ve x)$ and $h_a(\ve x)$, the multivariate CGP defined above provides a very flexible model and can model variant cross-correlation structures, and at the same time, can model the different correlation structure for each component. The covariance and cross-covariance at any two points $\vesub{x}{i}, \vesub{x}{j} \in \mathcal{R}^p$ can be calculated by
\begin{align}
\cov(\tau_a(\vesub{x}{i}),  \tau_a(\vesub{x}{j}))=&\cov(\xi_a(\vesub{x}{i}),  \xi_a(\vesub{x}{j}))+ \cov(\eta_a(\vesub{x}{i}),  \eta_a(\vesub{x}{j})), \nonumber \\
\cov(\tau_a(\vesub{x}{i}),  \tau_b(\vesub{x}{j}))=&\cov(\xi_a(\vesub{x}{i}),  \xi_b(\vesub{x}{j})), \ \mbox{ for } a, b =1, 2 \ (a \ne b). \label{cross}
\end{align}
If we take $h_{a}(\bs x)=v_{a0} \exp \{-\frac{1}{2} \bs{x}^{T}\bs{A}_{a0} \bs x \}$ and $g_a(\bs x)=v_{a1} \exp \{-\frac{1}{2} \bs{x}^{T}\bs{A}_{a1} \bs x \}$ for $a=1, 2$, the covariance in the first equation can be calculate by \eqref{42}, and the cross-covariance in the second equation is given by
\begin{displaymath}
\cov(\tau_a(\bs{x}_i),  \tau_b(\bs{x}_j))=(2 \pi)^{p/2} v_{10} v_{20} \lvert \bs{A}_{10}+\bs{A}_{20} \rvert ^{-1/2} \exp \{-\frac{1}{2} (\bs{x}_{i}-\bs{x}_{j})^T \bs{\ve \Sigma} (\bs{x}_{i}-\bs{x}_{j}) \},
\end{displaymath}
where $\ve \Sigma=\vesub{A}{10}(\vesub{A}{10}+\vesub{A}{20})^{-1}\vesub{A}{20}$.

Now let us look at the specific covariance structure of \eqref{43} using a discrete form. Consider
\begin{displaymath}
\bs{\tau} = \left \{ \tau_{1}(\bs{x}_{1i}),i=1,\ldots,n_{1}; \quad \tau_{2}(\bs{x}_{2j}) ,j=1,\ldots,n_{2} \right \},
\end{displaymath}
where $\bs{x}_{1i},\bs{x}_{2j} \in \mathcal{X} \subset \mathcal{R}^{p}$. Then $\bs\tau$ is a realization of a  multivariate CGP defined in \eqref{43}. It has an $(n_1+n_2)$-dimensional Gaussian distribution with zero means. Let $\bs{K}$ be the $(n_1+n_2) \times (n_1+n_2)$ covariance matrix of $\ve \tau$. It includes elements of $k_{ab}(\bs{x}_{ai}, \bs{x}_{bj})=\Cov(\tau_{a}(\bs{x}_{ai}),\tau_{b}(\bs{x}_{2j}))$ for $a, b \in \{1, 2\}$ and $i, j$ in either $\{1, \ldots, n_1\}$ or $\{1, \ldots, n_2\}$.

If we consider stationary processes, i.e. the covariance function depends only on the distance between two points
 $\bs{d}=\bs{x}_{ai}-\bs{x}_{bj}$, then the covariance function is defined by
\begin{equation}\label{subcov}
\begin{split}
k_{11}(\bs{d})&=k_{11}^{\xi_{1}}(\bs{d})+k_{11}^ {\eta_{1}}(\bs{d}) , \quad k_{12}(\bs{d})= k_{12}^{\xi_{12}}(\bs{d}),\\
k_{22}(\bs{d})&=k_{22}^{\xi_{2}}(\bs{d})+k_{22}^ {\eta_{2}}(\bs{d}) , \quad  k_{21}(\bs{d})= k_{12}^{\xi_{12}}(-\bs{d}),
\end{split}
\end{equation}
where, for example, $k_{12}^{\xi_{12}}(\bs{d})$ stands for the covariance between $\xi_1$ and $\xi_2$. It is straightforward to get the formulas if we use the squared exponential covariance function in  \eqref{42}. This can also be applied to other types of covariance functions.
We denote the multivariate GP defined above as a multivariate CGP (MCGP)
\begin{equation}
(\tau_1(\ve x), \tau_2(\ve x))^T \sim \textup{MCGP}(\xi_1(\ve x), \xi_2(\ve x), \eta_1(\ve x), \eta_2(\ve x)), \ \mbox{or} \ \textup{MGP}(0, k(\cdot, \cdot)), \label{mcgp}
\end{equation}
where $\xi_a$ and $\eta_a$ are defined in \eqref{e2} and \eqref{e1} respectively, and $\textup{MGP}(0, k(\cdot, \cdot))$ stands for a multivariate GP with zero mean and covariance function $k(\cdot, \cdot)$ which is determined by $\xi_a$ and $\eta_a$ in \eqref{subcov}. It is not difficult to extend the above bivariate case to a general multivariate case.

The covariance function defined by the above way is positive definite.

\textbf{Proposition 1.}\label{p1}
Assume that $\mathcal{S}(m)$ is an isotropic covariance function on $\mathcal{R}^{p}$, for any $p \in \mathbb{N}$.  If the function of covariance $ k_{ab}(\bs{d}) $ in \eqref{subcov} is given by
\begin{eqnarray*}
k_{ab}(\bs{d})=\frac{v_{a}v_{b}(2\pi)^{p/2}}{\mid \bs{A}_{a}+\bs{A}_{b}\mid^{1/2}}
\mathcal{S}(\sqrt {Q_{ab}(\bs{d};\bs{A}_a,\bs{A}_b)}),
\end{eqnarray*}
where
\begin{eqnarray*}
Q_{ab}(\bs{d};\bs{A}_a,\bs{A}_b)=\bs{d}^{T}\bs{A}_{a}(\bs{A}_{a}+\bs{A}_{b})^{-1}\bs{A}_{b}\bs{d}
\end{eqnarray*}
for any $v_{a}, v_{b} \in \mathcal{R}$ and arbitrary positive matrices $\bs{A}_{a}, a=1,2$, then the covariance function  defined in \eqref{subcov} is positive definite.

The proof is similar to the one given in \cite{Andriluka07} and the details can be found in \cite{Sofro16}.

For the squared exponential covariance function \eqref{42}, we have
\begin{equation}\label{46}
\begin{split}
k^{\xi_{a}}_{aa}(\bs{d})=&\frac{v_{a0}^{2}\pi^{p/2}} {\mid \bs{A}_{a0}\mid ^{1/2}} \exp \{- \frac{1}{2} Q_{aa}(\bs{d};\bs{A}_{a0},\bs{A}_{a0}) \},\\
k^{\xi_{ab}}_{ab}(\bs{d})=&
\frac{v_{a0}v_{b0}(2\pi)^{p/2}}{\mid \bs{A}_{a0}+\bs{A}_{b0}\mid ^{1/2}} \exp \{- \frac{1}{2}Q_{ab}(\bs{d};\bs{A}_{a0},\bs{A}_{b0}) \},\\
k^{ \eta_{a}}_{aa}(\bs{d})=&\frac{v_{a}^{2}\pi^{p/2}} {\mid \bs{A}_{a1}\mid ^{1/2}} \exp \{- \frac{1}{2} Q_{aa}(\bs{d};\bs{A}_{a1},\bs{A}_{a1}) \} \quad \mbox{ for } a,b=1,2 \mbox{ and } a \neq b .
\end{split}
\end{equation}

Similarly we can apply it to other covariance functions such as Matern and rational quadratics \citep{Shi11,Sofro16}.

\subsection{\label{}The Model}

Let $z_1$ and $z_2$ be two correlated response variables, for example the number of dengue fever and number of malaria cases in the example we discussed in Section 1. A general multivariate CGP model for dependent count data can be defined as follows.
\begin{equation}\label{eq51}
\begin{split}
z_{a}\mid \tau_{a} \sim & \textup{Poisson} (\mu_{a}),\\
\log(\mu_{a})=&\bs{U}_{a}^{T}\bs\beta_{a}+\tau_{a} (\vesub{x}{a}), \quad a=1,2,
\end{split}
\end{equation}
where $(\tau_1, \tau_2) \sim \textup{MCGP}(\xi_1, \xi_2, \eta_1, \eta_2)$, $\bs{U}_{a}$ is a set of covariates and a linear model is used here. Parametric $\vesub{\beta}{a}$ is used to describe the relationship between the response variable $z_a$ and the covariates $\bs{U}_{a}$.
The dependency of the observations for each component and the cross-correlation between components are modeled by $(\tau_1, \tau_2)$ via a MCGP. The cross-correlation or the cross-covariance is modeled by $\xi_1$ and $\xi_2$ in \eqref{e2}; while the covariance structure for each component is  modeled by $\xi_a$ and $\eta_a$. Since different covariance functions can be used for $\eta_a$ and $\xi_a$ for $a=1,2$, the model allows different covariance structures for each components. This largely increase the flexibility of the model, enabling the model to cope with data coming from different resources, having different data form and/or having different degrees of nonlinearity and smoothness. Model \eqref{eq51} uses MCGP to model multivariate Poisson data; for convenience, we call it as MCGP for Poisson data, or MCGPP in short.

In the above model,   $\bs{U}_{a}$ is a set of covariates to model the mean  while $\bs{x}_{a}$ is to model the covariance. Some of those covariates may be the same. In \eqref{eq51}, other parametric mean model can also be used. This will not add extra technical difficulty in the inference we will discuss next.

In model \eqref{eq51}, $\tau_a$ can  be treated as a nonlinear random effect. The posterior distribution can be calculated and the information consistency we will prove later in this section will guarantee  it approaches the underline true function if we have observations of sufficient large number.

Suppose that we have observed the following data
$\mathcal{D}=\{z_{ai},\bs{U}_{ai},\bs{x}_{ai}|a=1,2, ~i = 1,\ldots, n_a\}$,
where  $n_{1}$ and $n_{2}$ are the numbers of the observations for the two components respectively. Our model does not require the data is observed in pair, and those $n_1$ and $n_2$ could be different. Based on the model defined in \eqref{eq51}, $\bs{z}=(z_{11},\cdots,z_{1n_{1}},z_{21},\ldots,z_{2n_{2}})^{T}$ are conditional independent given $\ve \tau=(\bs{\tau}_{1}^T, \bs{\tau}_{2}^T)^T$, where $\vesub{\tau}{a}=(\tau_{a1}, \ldots, \tau_{an_a})^T$ for $a=1,2$. Thus,
\begin{equation} \label{pzt}
p(\ve z \mid \ve \tau) = \prod_{a=1}^2 \prod_{i=1}^{n_a} p(z_{ai} \mid \tau_{ai})
\end{equation}
where $p(z_{ai} \mid \tau_{ai})$ is the probability density of the Poisson distribution with mean $\vess{U}{ai}{T} \vesub{\beta}{a}+\tau_{ai}$.

Following the discussion in the last subsection, $\ve \tau$ is a realization of a MCGP. It has a $(n_1+n_2)$-dimensional Gaussian distribution with zero mean and covariance matrix $\ve K$. The element of $\ve K$ is calculated by equation \eqref{cross} and depends on the kernels $g_a$ and $h_a$ ($a=1,2$). Under a Bayesian framework, this defines a prior distribution of the latent variable $\ve \tau$. The related covariance functions  involve hyper-parameters, for example,  the squared exponential covariance function defined in  \eqref{46} depends on $\{v_{aj}, \vesub{A}{aj}, \ a=1, 2, j=0, 1 \}$. Although the values of those hyper-parameters (denoted by $\ve \theta$) can be given in advance based on prior knowledge, it is rather a difficult task if it is not impossible. This is because the physical meaning for some of them are not very clear, and the dimension of $\ve \theta$ is usually quite large. Among several different methods \citep{Shi11}, we adopt an empirical Bayesian approach in this paper, i.e. choosing the values of those hyper-parameters by maximising its marginal likelihood. Following the discussion in \citet{Wang14}, we can estimate $\bs\theta$ and other parameters, which are $\bs{\beta}_{a}$ in model \eqref{eq51}, at the same time.

\subsection{Estimation and prediction}

Given data $\mathcal{D}$, the marginal density of $\bs{z}$ given $\bs\beta$ and $\bs\theta$ is given by
\begin{eqnarray*}\label{55}
& p(\bs{z}\mid \bs\beta,\bs\theta,\bs{x})=\int p(\bs{z}\mid\bs \tau, \bs \beta)p(\bs\tau \mid\bs\theta)\textup{d}\bs\tau
=\int\left \{   \prod_{a=1}^{2}\prod_{i=1}^{n_{a}}p(z_{a_{i}}\mid \tau_{ai},\bs \beta_a)\right \}p(\bs\tau\mid\bs\theta)\textup{d}\bs\tau,
\end{eqnarray*}
and the marginal log-likelihood is
\begin{eqnarray}\label{57}
l(\bs \beta, \bs \theta)=\log\left \{p(\bs{z}\mid \bs \beta, \bs \theta,\bs{x})   \right \} =
\log \int \exp(\Phi(\bs\tau)) \textup{d} \bs\tau
\end{eqnarray}
where
\begin{eqnarray}\label{gpr9}
\Phi(\bs\tau)
=  -\frac{1}{2}\log\mid \ve{K}\mid-\frac{1}{2}\bs\tau^{T}\vesup{K}{-1}\bs\tau-\frac{n_{1}+n_{2}}{2}\log(2\pi) + \sum_{a=1}^{2} \sum_{i=1}^{n_{a}} \log [p(z_{a_{i}}\mid \tau_{ai},\bs{\beta}_a)],
\end{eqnarray}
with $\log p(z_{a_{i}}\mid \tau_{ai},\bs{\beta}_a)= z_{ai}\log(\mu_{ai})- \mu_{ai}- \log (z_{ai}!)$ and  $\mu_{ai}=\exp(\bs{U}_{ai}^{T}\bs\beta_{a}+\tau_{ai})$ for $a=1,2$. The integral involved in the above marginal likelihood is analytical intractable since the dimension of $\ve \tau$ is $n_1+n_2$, the total sample size, which is usually very large. We use a Laplace approximation. Let $\vesub{\tau}{0}$ be the maximiser of $\Phi(\ve \tau)$, we have
\begin{equation}\label{58}
\int \exp(\Phi(\bs\tau)) d \bs\tau \approx \exp\left \{\Phi(\bs\tau_{0})+\frac{n_{1}+n_{2}}{2}\log(2\pi)-\frac{1}{2}\log \mid\bs{H}\mid\right \}
\end{equation}
where $\bs{H}$ is the second derivative of $-\Phi(\ve \tau)$ respect to $\bs\tau$ and evaluated at $\bs\tau_{0}$.
Thus, $\bs{H}=\bs{C}+\vesup{K}{-1}(\bs\theta)$ and $\bs{C}$ is a diagonal matrix,
\begin{eqnarray*}
 \bs{C}&=& \diag \{\exp(\bs{U}_{11}^{T}\bs\beta_{1}+\tau_{011}),...,\exp(\bs{U}_{1n_{1}}^{T}\bs\beta_{1}+\tau_{01n_{1}}),\\
 &&\exp(\bs{U}_{21}^{T}\bs\beta_{2}+\tau_{021}),...,\exp(\bs{U}_{2n_{2}}^{T}\bs\beta_{2}+\tau_{02n_{2}}) \}.
 \end{eqnarray*}
We then estimate the parameters by  maximising the likelihood function with Laplace approximation in equation \eqref{58}.





We now turn to calculate  prediction of $\bs{z}^{*} = (z_{1}^{*},z_{2}^{*})^{T}$ at a new point with  $\bs{U}^{*} = (\bs{U}_{1}^{*}, \bs{U}_{2}^{*})$ and $\bs{x}^*=(\bs{x}_{1}^{*},\bs{x}_{2}^{*})$. We still use $\mathcal D$ to denote all the training data and assume that the model itself has been trained (all unknown parameters have been estimated).  We will calculate the predictive mean $E({\bs{z}^{*}}\mid \mathcal D)$ as well as the predictive variance $\Var(\ve z^* \mid\mathcal D)$.

Let $\bs\tau^{*}=\bs\tau(\bs{x}^{*})=(\tau_{1}^{*},\tau_{2}^{*})^T$ be the underlying latent variable at $\bs{x}^{*}$.  The expectation of $\bs{z}^{*}$ conditional on $\bs\tau^{*}$ is  given by
\begin{eqnarray*}
\E({\bs{z}^{*}}\mid\bs\tau^{*},\mathcal D)= \begin{pmatrix}
  \E({z_{1}^{*}}\mid\tau_{1}^{*},\mathcal D)\\
    \E({z_{2}^{*}}\mid\tau_{2}^{*},\mathcal D)
                   \end{pmatrix}=\begin{pmatrix}
 \exp(\bs{U}_1^{*T}\hat{\bs\beta}_1+\tau_{1}^{*})\\
    \exp(\bs{U}_2^{*T}\hat{\bs\beta}_2+\tau_{2}^{*})
                  \end{pmatrix}\triangleq \exp (\bs{U}^{*T} \hat {\bs{\beta}}+\bs \tau^{*}).
\end{eqnarray*}
It follows that
\begin{equation}\label{59}
\E(\bs{z}^{*}\mid \mathcal D)=\E[\E(\bs{z}^{*} \mid \bs \tau^{*},\mathcal D)]=\int
 \exp (\bs{U}^{*T} \hat {\bs{\beta}}+\bs \tau^{*})p(\bs \tau^{*}\mid \mathcal D) \textup{d}\bs \tau^{*}.
\end{equation}
Note that
\begin{align}\label{gpr11a2}
p(\bs\tau^{*}\mid \mathcal D)
&=\int p(\bs\tau^{*}\mid \bs\tau,\mathcal D)p(\bs\tau \mid \mathcal D)\textup{d}\bs\tau \nonumber \\
&=\int  p(\bs\tau^{*},\bs\tau \mid \mathcal D)\textup{d}\bs\tau
= \frac{1}{p(\bs{z})}\int p(\bs{z}\mid \bs\tau) p(\bs\tau^{*},\bs\tau)\textup{d}\bs\tau .
\end{align}
Hence, equation \eqref{59}  above can be rewritten as
\begin{equation}\label{512}
\E({\bs{z}^{*}}\mid  \mathcal D)=\frac{1}{p(\bs{z})}\int \int \exp (\bs{U}^{*T} \hat {\bs{\beta}}+\bs \tau^{*}) p(\bs{z}\mid\bs\tau) p(\bs\tau^{*},\bs\tau)\textup{d}\bs\tau \textup{d}\bs\tau^{*}.
\end{equation}
For convenience we denote $\vesub{\tau}{+}=(\bs{\tau}^T,\bs{\tau}^{*T})^{T}$, which is a realization of the MCGPP defined in \eqref{eq51}. So its density function is a multivariate normal distribution with zero mean. The  ${(n_{1}+n_{2}+2) \times (n_{1}+n_{2}+2})$  covariance matrix is calculated similar to $\ve K$ in \eqref{gpr9}, and it is denoted by $\vesub{K}{+}$.  Thus, the  above equation  can be written as
\begin{eqnarray}
\E({\bs{z}^{*}}\mid \mathcal D)
&=&\frac{1}{p(\bs{z})}\int \exp (\bs{U}^{*T} \hat {\bs{\beta}}+\bs \tau^{*})
 [p(\vesub{z}{1} \mid \hvesub{\beta}{1}, \vesub{\tau}{1})  p(\vesub{z}{2} \mid \hvesub{\beta}{2}, \vesub{\tau}{2})  ]  \nonumber \\
&& \ \ \ \ \ \ \left [  (2\pi)^{-\frac{(n_{1}+n_{2}+2)}{2}} \mid\ve{K}_{+}\mid^{-\frac{1}{2}} \exp (-\frac{1}{2}\bs\tau_{+}^{T}\ve{K}_{+}^{-1}\bs\tau_{+})\right ] \textup{d}\ve \tau_{+} \nonumber \\
&=&\frac{1}{p(\bs{z})}\int \exp ({\tilde{\bs\Phi}(\bs\tau_{+})} )\textup{d} \bs\tau_{+}. \label{513}
\end{eqnarray}
where
\begin{eqnarray}
\tilde{\bs\Phi}(\bs\tau_{+})
&=& \bs{U}^{*T} \hat {\bs{\beta}}+\bs \tau^{*}+\sum_{i=1}^{n_{1}}\log p(z_{1i}\mid\hat{\bs\beta}_{1},\tau_{1i})+\sum_{i=1}^{n_{2}}\log p(z_{2i}\mid\hat{\bs\beta}_{2},\tau_{2i}) \nonumber \\
&& -\frac{n_{1}+n_{2}+2}{2}\log(2\pi)-
\frac{1}{2}\log \mid \ve{K}_{+}\mid -\frac{1}{2}\bs\tau_{+}^{T}\ve{K}_{+}^{-1}\bs\tau_{+}, \label{phits}
\end{eqnarray}
where $p({z}_{ai}\mid\bs\beta_{a},\bs\tau_{ai})$ is the density of the Poisson distribution with mean  $\mu_{ai}=\exp(\bs{U}_{ai}^{T}\bs\beta_{a}+\tau_{ai})$ for $a=1, 2$.
The calculation of the integral is difficult and we also use a Laplace approximation:
\begin{eqnarray}\label{514}
\int \exp (\tilde{\bs\Phi}(\bs\tau_{+})) \textup{d} \bs\tau_{+} \approx \exp \{ \tilde{\bs\Phi}(\hat{\bs\tau}_{+})+\frac{n_{1}+n_{2}+2}{2}\log(2\pi)
-\frac{1}{2}\log \mid\ve{K}_{+}^{-1}+\widehat{\bs{C}}_{+}\mid \}
\end{eqnarray}
where $\widehat{\bs{C}}_{+}$ is  the second derivative of the first four items in \eqref{phits}
with respect to $\bs\tau_{+}$ and evaluated at $\hat{\bs\tau}_{+}$. It is an $(n_1+n_2+2)$ dimensional  diagonal matrix:
\begin{eqnarray*}
 \widehat{\bs{C}}_{+}&=&\diag (\exp(\bs{U}_{11}^{T}\hat{\bs\beta}_{1}+\hat{\tau}_{11}),...,\exp(\bs{U}_{1n_{1}}^{T}\hat{\bs\beta}_{1}+\hat{\tau}_{1n_{1}}),\\
&& \exp(\bs{U}_{21}^{T}\hat{\bs\beta}_{2}+\hat{\tau}_{21}),...,
 \exp(\bs{U}_{2n_{2}}^{T}\hat{\bs\beta}_{2}+\hat{\tau}_{2n_{2}}), 0, 0 ).
 \end{eqnarray*}






Similarly, we can calculate the predictive variance, which is defined as
\begin{equation}
\Var(\bs{z}^{*}\mid\mathcal D)=\begin{pmatrix}
                        \Var(z_{1}^{*}\mid\mathcal D) & \Cov(z_{1}^{*},z_{2}^{*}\mid\mathcal D)\\
                        \Cov(z_{1}^{*},z_{2}^{*}\mid\mathcal D)& \Var(z_{2}^{*}\mid\mathcal D)
                        \end{pmatrix},
\end{equation}
where
\begin{eqnarray}\label{516}
\Var({z}^{*}\mid \mathcal D)&=&\E[\Var({z}^{*}\mid \bs \tau^{*},\mathcal D)]+\Var[\E({z}^{*}\mid\bs\tau^{*},\mathcal D)
\end{eqnarray}
Here $z$ could be either $z_1$ or $z_2$.
Because $\Var({z}^{*}\mid\bs\tau^{*},\mathcal D)=\E({z}^{*}\mid\bs\tau^{*},\mathcal D)$ for a Poisson distribution, we have
$\E[\Var({z}^{*}\mid\bs\tau^{*},\mathcal D)]=\E({z}^{*}\mid\mathcal D)$. The second item can be calculated by
\begin{eqnarray}\label{517}
\Var[\E(\bs{z}^{*}\mid\bs\tau^{*},\mathcal D)]&=&\E[\E(\bs{z}^{*}\mid \bs\tau^{*},\mathcal D)]^{2}-[\E[\E(\bs{z}^{*}\mid\bs\tau^{*},\mathcal D)]]^{2}\nonumber\\
&=&\int (\exp (\bs{U}^{*T} \hat {\bs{\beta}}+\bs \tau^{*}))^{2}p(\bs \tau^{*}\mid \mathcal{D}) \textup{d}\bs \tau^{*}-[\E(\bs{z}^{*}\mid \mathcal D)]^{2}.
\end{eqnarray}
The first item in \eqref{517} can be obtained by Laplace approximation using the similar way  to calculate $\E(\bs{z}^{*}\mid \mathcal D)$ in \eqref{513}.

The covariance $\Cov(z_{1}^{*},z_{2}^{*}\mid\mathcal D)$ is calculated by
\begin{eqnarray}\label{518}
\Cov(z_{1}^{*},z_{2}^{*}\mid\mathcal D)&=&\E[z_{1}^{*} z_{2}^{*}\mid\mathcal D]-\E[(z_{1}^{*}\mid\mathcal D)]\E[(z_{2}^{*}\mid\mathcal D)] \nonumber \\
&=&\E \{\E[z_{1}^{*} z_{2}^{*}\mid \vesup{\tau}{*}, \mathcal D]\}-\E[(z_{1}^{*}\mid\mathcal D)]\E[(z_{2}^{*}\mid\mathcal D)].
\end{eqnarray}
The  first item in \eqref{518} is similar to the first item in \eqref{517}, and can be calculated by Laplace approximation.
\subsection{\label{}Consistency}

The prediction based on a {GPR} model is consistent when the sample size of the data collected from a certain curve is sufficiently large and the covariance function satisfies certain regularity conditions \citep{cho,see}. The consistency does not depend on the common mean structure or the choice of the values of hyper-parameters involved in the covariance function.

In this section, we will discuss information consistency and extend it to a more general context than the result of \cite{Wang14}.  We focus on $\tilde{\bs{z}}$ to $\bs{z}$, where $\tilde{\bs{z}}=(\tilde{z}_{11},...\tilde{z}_{1n_{1}},\tilde{z}_{21},...,\tilde{z}_{2n_{2}})$ are predicted observations and $\bs{z}=(z_{11},...,z_{1n_{1}},z_{21},...,z_{2n_{2}})$ are actual observations, and $n_{1}$ and $n_{2}$ are the number of observations of the first input and the second input respectively.  The corresponding covariate are $\bs{X}_{n_{1}n_{2}}=\{(\bs{x}_{1i},\bs{x}_{{2j}}), i=1, \ldots, n_1, j=2, \ldots, n_2 \}$ where $\bs{x}_{{ai}} \in \mathcal{X} \subset \mathbb{R}^{p}$  are independently drawn from its distribution, and the latent variable is  $(\tau_{{1i}},\tau_{{2j}})$.

We assume that $z_{{1i}}$ and $z_{{2j}}$ is a set of samples and follow a  bivariate Poisson distribution with $\mu_{1i}=\exp(\bs{U}_{1i}^{T}\bs \beta_{1}+ \tau_{1i}(\bs{x}_{1i}) )$  and $ \mu_{2j}=\exp(\bs{U}_{2j}^{T}\bs \beta_{2}+ \tau_{2j}(\bs{x}_{2j}))$ respectively and $ ({\tau}_{1i}(\mathord{\cdot}),\tau_{2j}(\mathord{\cdot})) \sim  \textup{MGP} (\bs{0},{k}(\mathord{\cdot},\mathord{\cdot}))$ was discussed in the previous section. Therefore, the stochastic process $\tau_{1}(\mathord{\cdot})$ and $\tau_{2}(\mathord{\cdot})$ induces a measure on space $\mathcal{F}:\left \{f(\mathord{\cdot}): \mathcal{X}\rightarrow \mathbb{R}  \right \}
$. For convenience, we can rewrite
$
\bs{z}=(z_{11},...,z_{1n_{1}},z_{21},...,z_{2n_{2}})
=(z_{1},...,z_{n_{1}},z_{n_{1}+1},...,z_{n_{1}+n_{2}})
$
and the covariate as $\bs{X}_{n_{1}n_{2}}=(\bs{x}_{1},...,\bs{x}_{n_{1}},\bs{x}_{n_{1}+1},...,\bs{x}_{n_{1}+n_{2}})$. Let
$\mathcal{D}_{n_{1}n_{2}}=\left \{ (\bs{x}_{i},z_{i}), i=1,..., n_{1}+n_{2} \right \}$, we have 
\begin{eqnarray*}
\E({\bs{z}}|\bs\tau) \triangleq\exp (\bs{U}^{T}\hat {\bs{\beta}}+\bs{\tau}(\bs{x})).
\end{eqnarray*}
Suppose that the hyper-parameters $\bs \theta$ in the covariance function are estimated by an empirical Bayesian method and the estimator is denoted by $\tilde {\bs \theta}$. Let $ \tau_{0}$  be the true underlying function, i.e. the true mean of  $z_{i}$ is given by $\mu_{i0}=\exp(\bs{U}_{i}^{T}\bs\beta+\tau_{0}(\bs{x}_{i}))$.
Denote
\[p_{mgp}(\bs{z})= \int p(z_{1},...,z_{n_{1}},z_{n_{1}+1},...,z_{n_{1}+n_{2}}| \ve \tau(\ve x)) p_{n_1+n_2}( \ve \tau)\textup{d}\ve \tau\]
and
\[p_{0}(\bs{z})=p(z_{1},...,z_{n_{1}},z_{n_{1}+1},...,z_{n_{1}+n_{2}}| \tau_{0}(\ve x)),\]
then $p_{mgp}(\bs{z})$ is the Bayesian predictive distribution of $\bs{z}$ based on a MCGPP model. Note that $p_{n_1+n_2}( \ve \tau)$ depends on the sample size $n_{1}+n_{2}$ since the hyper-parameters of $\ve \tau$ are estimated from the data. We say that $p_{mgp}$ achieves information consistency if
\begin{equation}\label{}
\frac{1}{n_{1}+n_{2}}\E_{\bs{X}_{n_{1}n_{2}}}\left ( D[p_{0}(\bs{z}),p_{mgp}(\bs{z})] \right )\rightarrow 0 \quad as \quad n_{1} \rightarrow \infty \quad and \quad n_{2} \rightarrow \infty,
\end{equation}
where $\E_{\bs{X}_{n_{1}n_{2}}}$ denotes the expectation under the distribution of $\bs{X}_{n_{1}n_{2}}$ and $ D[p_{0}(\bs{z}),p_{mgp}(\bs{z})]$ is the Kullback-Leibler divergence between $p_{0}(\cdot)$ and $p_{mgp}(\cdot)$, i.e.,
\[D[p_{0}(\bs{z}),p_{mgp}(\bs{z})]=\int p_{0}(\bs{z}) \log \frac{p_{0}(\bs{z})}{p_{mgp}(\bs{z})} d\bs{z}.\]

\textbf{Theorem 1.} \label{t1}
Under the MCGPP model \eqref{eq51} and the condition given in Lemma 1 in Appendix, the prediction $\hat{\bs{z}}$ is information consistent  if the RKHS norm $\left \| \tau_{0} \right \|^{2}_{\bs{K}_{n_{1}n_{2}}}$ is bounded and the expected regret term $\E_{\bs{X}_{n_{1}n_{2}}}(\log |\bs{I}+\delta\bs{K}_{n_{1}n_{2}} |)=o(n_{1}+n_{2})$. The error bound is specified in \eqref{con4}.

The proof of the theorem is  given in Appendix.

\textbf{Remark 1}
The regret term $R=\log|\bs{I}+\delta\bs{K}_{n_{1}n_{2}}|$ depends on the covariance function $k(\bs{x}_{i},\bs{x}_{j})$ for a convolved bivariate GP and the distribution of $\bs{x}$.   We can use it to identify the upper bounds of the expected regret for some commonly used covariance functions by extending results in \cite{Wang14}.  The detailed discussion is given in Appendix.

\section{\label{}Numerical Results}
In this section, we demonstrate the performance of the proposed method by comprehensive simulation studies with two scenarios and also present results for two real data examples.
\subsection{\label{}Simulation Studies: Scenario 1}
In the first scenario, we use a discrete bivariate Poisson regression model in (\ref{eq51}) as the true model to generate data:
\begin{eqnarray}\label{531a}
\begin{pmatrix}
 z_{1i}(\bs{x}_{i})\\
 z_{2j}(\bs{x}_{j})
\end{pmatrix} \sim \begin{pmatrix}
               \textup{Poisson} (\mu_{1i}(\bs{x}_{i})),\quad i=1,\ldots, n_{1}\\
                \textup{Poisson} (\mu_{2j}(\bs{x}_{j})),\quad j=1,\ldots,n_{2}
               \end{pmatrix},
\end{eqnarray}
where
\begin{eqnarray}
\begin{pmatrix}
  \mu_{1i}(\bs{x}_{i})=\exp(\bs{U}_{1i}^{T}\bs{\beta}_{1}+ \tau_{1i}(\bs{x}_{i}) )\nonumber\\
  \mu_{2j}(\bs{x}_{j})=\exp(\bs{U}_{2i}^{T}\bs{\beta}_{2}+ \tau_{2j}(\bs{x}_{j}))
 \end{pmatrix}, \quad  \quad  \begin{pmatrix}
\tau_{1i}(\mathord{\cdot})\\
 \tau_{2j}(\mathord{\cdot})
\end{pmatrix} \sim \textup{MGP}(\bs{0},{k}(\mathord{\cdot},\mathord{\cdot})),
\end{eqnarray}
and ${k}(\mathord{\cdot},\mathord{\cdot})$ is defined by \eqref{cross}  and \eqref{subcov}. We take   $\beta_{10}=1$, $\beta_{11}=2$, $\beta_{20}=1$ and $\beta_{21}=2$.

Random processes $\tau_{1i}$ and $\tau_{2i}$ are generated from a MGP with a mixed covariance structure, the combination of two different covariance functions.   Specifically, $\eta_1$ is generated from  a GP with the squared exponential covariance function with $v_{11}=0.04$ and $A_{11}=1$, while $\eta_2$ from the Gamma exponential covariance function with $v_{21}=0.04$ and $A_{21}=1$. The shared processes $ \xi_a$'s follow the squared exponential covariance function with $v_{10}=0.04, v_{20}=0.04, A_{10}=1$ and $A_{20}=1$. The covariates $\bs{x}_{i}$'s are equally spaced in $[-5,5]$. Recall that $\tau_a=\xi_a+\eta_a$ for $a=1,2$. Thus  $\bs \tau=\left \{ \tau_{1i},\tau_{2j} \right \}$ is dependent GPs but have different covariance structure for each component. We set $n_1=n_2=20$.

As we discussed in the previous section, the proposed MCGPP model allows different covariance structure for each component and thus it should be able to have a good fit for the data generated using the above way. To show the stability of the models, we considered the model \eqref{eq51} with the following covariance functions.\\
 \textit{Model 1}~--~$\xi_{1}, \xi_{2}$ and $\eta_{1}$ have  squared exponential covariance functions and $\eta_{2}$ has a Gamma exponential  covariance function, i.e. this model assumes the same covariance structure as the true model;  \\
\textit{Model 2}~--~all $\eta_{1},\eta_{2},\xi_{1}$ and $\xi_{2}$ have rational quadratic covariance functions; \\
\textit{Model 3}~--~all $\eta_{1},\eta_{2},\xi_{1}$ and $\xi_{2}$ have Matern covariance functions; \\ \textit{Model 4}~--~all $\eta_{1},\eta_{2},\xi_{1}$ and $\xi_{2}$ have squared exponential covariance functions. \\
 As comparison, we also consider the model in \cite{Crainiceanu12} ({CDR}), where  $\bs{\tau}_{1}$ is a GP with zero mean and a squared exponential covariance function, and $\bs{\tau}_{2}$ is conditional on $\tau_1$, i.e. 
$\bs{\tau}_{2}\mid \bs{\tau}_{1} \sim  \mathcal{N} (\alpha \bs{\tau}_{1}, \sigma_{\epsilon}^{2})$.
The dependency is determined by $\alpha$. It is a useful model but lack of flexibility on modelling covariance structures for multiple components since the covariance structure of the second component is determined by the first one.

We also compared them to the independent model ({Indep}).  In this case, we assume that $\tau_1$ and $\tau_2$ are independent and each follows a GP with a squared exponential covariance function.

We use each of the six models to fit the data. To measure the performances of those models, we further generate a new set of test data (20 for each component) and use the fitted model to calculate the prediction of $\mu_{ai}, \ a=1,2$ and $i=1, \ldots, 20$ for the test data. We  then calculate the root mean squared error (RMSE) between the predictions and the test data for $\mu_{ai}$. Table \ref{5predr3} listed the average RMSEs based on 100 replications. As expected, Model 1 gives the best result. Models 2 to 4 also give reasonably good results although different covariance functions are used in those models. This shows that the proposed model is flexible to fit  data with different covariance structure in each component, and is robust as well. Model CDR models the dependency using a conditional distribution, i.e. the covariance structure of the second component is dependent on the first one. When this model is applied to the data having different covariance structures for each component, the result is not satisfactory. Model Indep ignores the dependency between components and consequently has large errors.

\begin{table}[h!]
	\centering
		\caption{ Average RMSEs  between $\mu$ and $\hat{\mu}$ based on one hundred replications.}\label{5predr3}
	\begin{tabular}{cc}
		\toprule
		Model & Average RMSE \\
		\midrule
		 {Model 1} & 0.02627\\
		 {Model 2} & 0.03841 \\
		 {Model 3} & 0.03028 \\
		 {Model 4} & 0.03459\\
		 {CDR} & 0.10920\\
		{Indep} & 0.04628\\
		\bottomrule
	\end{tabular}%

\end{table}%

We also calculate the difference between the estimation of $\beta$ and its true values. The values of RMSE between $\hat \beta $ and its true value and the sampling bias based on 100 replications are presented in Table \ref{512}. The findings are almost the same as those from Table \ref{5predr3}.

\begin{table}[h!]
\centering
\small
\caption{RMSEs between $\hve \beta$  and their true values and the absolute value of the sampling bias (in parenthesis)  based on one hundred  replications.}\label{512}
\begin{tabular}{ccccc}
	
\toprule
 &\multicolumn{3}{c}{RMSE ($|$bias$|$)}\\
\cmidrule(lr){2-5}
    Model &$\beta_{11}$ &$\beta_{12}$  &$\beta_{21}$ & $\beta_{22}$\\
\midrule
{Model 1} & 0.03496 (.000)  & 0.04547 (.004)   &  0.03967 (.005) & 0.03739 (.007)\\
{Model 2} &0.03381 (.003) & 0.04130 (.000)   &  0.03802 (.001)& 0.03626(.004) \\
{Model 3} & 0.03478 (.005) & 0.05156 (.002)   &  0.04036 (.007) & 0.03279 (.000) \\
{Model 4} & 0.04833 (.003) & 0.04560 (.001)  &  0.04106 (.004)  & 0.04066 (.000)\\
{CDR} & 0.13076  (.020) & 0.17025 (.026)  &  0.13972 (.017) & 0.15640 (.005)\\
{Indep} & 0.09486 (.009) & 0.13251  (.018)  & 0.14912  (.022) & 0.11417 (.013)\\
\bottomrule
\end{tabular}

\end{table}

\subsection{\label{}Simulation Studies: Scenario 2}

We now consider a scenario with multidimensional covariates and nonlinear mean function. The model is define as
\begin{eqnarray*}
	\bs \mu_{1i}(\bs{x}_{i})&=& \exp(\bs{y}_{1i}(\bs{x}_{i})),\quad \bs{z}_{1i}(\bs{x}_{i})\sim  \textup{Poisson}(\bs \mu_{1i}(\bs{x}_{i})), \quad i=1,\ldots,n_{1},\\
	\bs \mu_{2j}(\bs{x}_{j})&=& \exp(\bs{y}_{2j}(\bs{x}_{j})),\quad
	\bs{z}_{2j}(\bs{x}_{j})\sim  \textup{Poisson}(\bs \mu_{2j}(\bs{x}_{j})), \quad j=1,\ldots,n_{2},
\end{eqnarray*}
The latent variables $y_{1i}(\vesub{x}{i})$ and $y_{1j}(\vesub{x}{j})$ are generated by the following way
\begin{eqnarray*}
y_{1i}(\bs{x}_{i})&=& 0.2 {x}_{1i} \cdot  \mid {x}_{1i}\mid^\frac{1}{3}+\log ({x}_{2i})+\tau_{1i}(\bs{x}_{i}),\quad i=1,\ldots,n_{1},\\
y_{2j}(\bs{x}_{j})&=& sin ({x}_{2j})+0.4 {x}_{2j} \cdot  \mid{x}_{1j}\mid^\frac{1}{4}+\tau_{2j}(\bs{x}_{j}),\quad j=1,\ldots,n_{2},
\end{eqnarray*}
where $(\tau_{1i}(\mathord{\cdot}), \tau_{2j}(\mathord{\cdot})) \sim {MGP}(\bs{0},{k}(\mathord{\cdot},\mathord{\cdot}))$ and ${k}(\mathord{\cdot},\mathord{\cdot})$ is the same as the one in Scenario 1.
$\bs{x}=\left \{{x}_{1i},{x}_{2j}  \right \}$ are equally spaced in $[-5,10]$ and $[1,2]$ respectively and $\bs \tau=\left \{ \tau_{1i},\tau_{2j} \right \}$ is dependent GP which is formed in the same way to Scenario 1 in Model 1,  i.e.\  a mixed squared exponential covariance function and a Gamma exponential covariance function. Also the true values are the same as those used in Scenario 1.

In each replication, we generate $n_1=n_2=20$ observations as training data, and the further same numbers of observations as test data. We used all six models defined in Scenario~1 to fit the data. Bear in mind that, although we assumed the same covariance structures in Model 1  as those in the true model, Model 1 is different to the true model since nonlinear mean model is used in the true model while only linear mean model is assumed in the proposed model (i.e. Models 1 to 4).  \cite{Shi12} argued that the GPR is a flexible nonlinear Bayesian model and can fit nonlinear curves for continuous Gaussian data. We expect Models 1 to 4 can also fit the nonlinear latent curves, and thus they should provide a good fit to the non-Gaussian Poisson data in this scenario. The simulation study results presented in Table \ref{53} confirm the expectation.  The numbers in the table is the average RMSE between the generated value of $\mu$ and its prediction $\hat \mu$ based on 100 replications. The very small values of RMSE indicate that that GPR model is good on fitting the nonlinear data.

 Different covariance functions are used in Models 2 to 4, but all of them provide reasonable good results and all are better than Models CDR and Indep, where CDR models the covariance structure by a conditional approach, and Model Indep assumed independence between two components.

\begin{table}[h!]
  \centering
   \caption{ The average RMSE between $\mu$ and $\hat{\mu}$ based on one hundred replications.}\label{53}
      \begin{tabular}{cc}
    \toprule
      Model &Average RMSE  \\
    \midrule
     {Model 1} & 0.020587\\
     {Model 2} & 0.022521 \\
     {Model 3} & 0.022453 \\
     {Model 4} & 0.023001\\
     {CDR} &0.028196\\
     {Indep} &0.025159\\
    \bottomrule
    \end{tabular}%
 \end{table}%

\subsection{\label{}Real Data Analysis}
We will present results for two real sets of data. The first one is  data relating to two type of cancers  in Minnesota, USA. The second data concern Dengue fever and Malaria in Indonesia.
\noindent {\bf 1. Lung and Oesophageal Cancer data}

From information on the NHS web site (www.nhs.uk), one of the most dangerous and common types of cancer is lung cancer. Every year there are around  44,500 people diagnosed with this condition. The symptoms usually do not always appear in the early stages, although  some symptoms develop in many people, such as blood or persistent coughing, breathlessness and weight loss.
In over 85 percent of cases, the main cause of lung cancer is cigarette smoking  although people who have never smoked can be diagnosed with this cancer.  Smoking  can cause other cancers, such as  oesophageal cancer and mouth cancer.

There are more than 8,500 new cases of  oesophageal cancer diagnosed each year in the UK which means that this cancer is uncommon but is not rare. As with lung cancer, smoking and drinking alcohol are the highest risk factors for this cancer.

Fig 1 in \cite{Jin05} present the number of cases for each cancer in Minnesota,  USA. The map   shows clearly that the county-level maps of the age-adjusted standardized mortality ratios  between lung and oesophageal have a positive correlation across region or area. Thus it is better to investigate those two cancers using a joint multivariate model.

\cite{Jin05} analysed the relationship between lung cancer and oesophageal cancer using a generalized intrinsics autoregressive model which  was based on neighbourhood for each region as the main effect of the model. In practice, this model may have  difficulty in prediction  due to problem of defining  the neighbourhood for each area. Similar to   {CDR} model in \citep{Crainiceanu12}, a conditional approach is used in \cite{Jin05} to define the cross-correlation between two components which is a less flexible model as we discussed in simulation studies.

We  use MCGPP model here.
The model can be written as
\begin{equation}
z_{ia} \sim \textup{Poisson} (E_{ia}e^{\tau_{ia}(\bs{x}_{i})}), \quad i=1,...,87, \quad a=1,2,
\end{equation}
where $z_{ia}$ is the observed number of deaths due to cancer $a$ in county $i$, $E_{ia}$ is the corresponding expected number of deaths (assumed known) and
$\tau_{ia}(\mathord{\cdot}) \sim {MGP}(\bs{0},{k}(\mathord{\cdot},\mathord{\cdot}))$ which is explained in equation  \eqref{subcov}. Here, $\bs{x}$ are defined from spaced point values of latitude and longitude, the location, of each county. The correlation of the mortalities between two areas depends on their locations. The nearer, the larger. This is similar to the assumptions in \cite{Jin05}, but it is straightforward to find the values of $\ve x$, and the covariance structure can be learned and adjusted from the data  in MCGPP model.

As a comparison,   we also used  {CDR} model.

To measure the performance, we select data randomly from the whole data set to form training data consisting of two thirds of the data and the remainder is used for test data. We estimate parameters by an empirical Bayesian approach using the training data and then calculate prediction for the test data, and the  value of  error rate between the predictions and the actual observations.    Table \ref{c6} reports the  average ERs based on ten replications.  It shows that the MCGPP model provides very accurate results and is better than CDR. AIC (using the full data) also support the MCGPP model.
\begin{table}[h!]
\centering
\small
\caption{Numerical results for cancer data}\label{c6}
\begin{tabular}{ccc}
\hline
Method&  Average ER & AIC \\
\hline
CDR &  0.0149 & 1640.202\\
MCGPP     &0.0080 &1399.822 \\
\hline
\end{tabular}

\end{table}

\noindent {\bf 2. Dengue Fever and Malaria data }

We now analyse dengue fever and malaria data in Indonesia. Both of the diseases can be spread by two different types mosquitoes which are hard to distinguish from each other.  Therefore, it is more sensible to analyse them together in a joint multivariate model. The data are also  spatially correlated. We compared several methods to deal with this spatial effect, including MCGPP,   an intrinsic autoregressive model (CAR), and a conventional Poisson regression model. Among all those models, we found MCGPP are the best for the data; the details can be found in  \citep{Sofro16}.

We present three  models here taken from the different set of multidimensional covariates used in modelling covariance structure in MCGPP. The first model involves location (latitude and longitude) and all five observed covariates (health water ($x_{1}$), healthy  rubbish bin ($x_{2}$), waste water disposal facilities ($x_{3}$), clean and healthy behaviour ($x_{4}$) and healthy house ($x_{5}$)).  The second  model uses location and three  covariates, $x_{1} , x_{2} , x_{3}$. The last model uses the location only.
\begin{table}[h!]
\centering
\small
\caption{The average of error rate  based on fifteen replications}\label{c53}
\begin{tabular}{ccc}
\toprule
&\multicolumn{2}{c}{Average ER}\\
\cmidrule(l){2-3}
Models& MCGPP& CDR\\
\midrule
Full (location and all covariates)    &  0.000994  & 0.001374\\
Location and $x_{1}, x_{2}, x_{3}$ & 0.001018  & 0.002000\\
Location                                         & 0.001137  & 0.002252 \\
\bottomrule
\end{tabular}

\end{table}

Similar to the previous example, we also calculate the error rate for the test data. The results based on fifteen replications are presented in  Table \ref{c53}. Not surprisingly, the first model provides the best result. However, the second model performs almost as well as the first one, indicating $x_1$, $x_2$ and $x_3$ are the most important facts related to both diseases. As a comparison, we also present the results by using  CDR model. It gives less accurate results.

\section{Conclusions}

In this paper, we proposed a new method for multivariate Poisson regression analysis for dependent count data using convolved Gaussian processes. It is a very flexible model, can model nonlinear data, allow different covariance structure for each component, and also copy with multidimensional covariates. The approach is also quite robust, providing reliable results even when different covariance functions are used.

We limited our discussion in this paper to the bivariate case, the idea can be used to general multivariate cases. However, it is worth a further investigation on how to define cross-correlation for multiple components and how to implement the method efficiently.

\newpage
\section*{Appendix : Proof of information consistency}\label{5ap}
The proof presented below is an extension from consistency theorem in \cite{Wang14}.
\newline
\textbf{Lemma 1}\label{l1}\newline
Suppose $z_{1i}$ and $z_{2j}$  are conditional independent samples from a bivariate Poisson distribution given \eqref{eq51} and $\tau_{0} \in \mathcal{F}$ has a multivariate convolved Gaussian prior with zero mean and bounded covariance function $k(\cdot,\cdot)$ for any covariate values in $\mathcal{X}$. Suppose that $k(\cdot,\cdot)$ is continuous in $\bs \theta$ and the estimator $\hat{\bs\theta}\rightarrow \bs \theta$ almost surely as $n_{1}\rightarrow \infty $ and  $n_{2}\rightarrow \infty $. Then
\begin{eqnarray}\label{2}
&&-\log p_{mgp}(z_{1},...,z_{n_{1}+n_{2}})+\log p_{0}(z_{1},...,z_{n_{1}+n_{2}})  \nonumber\\
 \leqslant &&\frac{1}{2}\left \| \tau_{0} \right \|^{2}_{\bs{K}_{n_{1}n_{2}}} +\frac{1}{2}  \log|\bs{I}+\delta\bs{K}_{n_{1}n_{2}}| + C
\end{eqnarray}
where $\left \| \tau_{0} \right \|^{2}_{\bs{K}_{n_{1}n_{2}}}$ is the reproducing kernel Hilbert space (RKHS) norm of $\tau_{0}$ associated with $k(\cdot,\cdot)$, $\bs{K}_{n_{1}n_{2}}$ is the covariance matrix of $\tau_{0}$ over the covariate $\bs{X}_{n_{1}n_{2}}$, $\bs{I}$ is the $(n_{1}+n_{2})\times( n_{1}+n_{2})$ identity matrix, $\delta$ and $C$ are some positive constants.

\begin{proof}
In this proof, we use a covariance function to define a function on $\mathcal{X}$. The space of such a function is known as a reproducing kernel Hilbert space (RKHS) . Let $\mathcal H$ be RKHS associated with covariance function $k(\cdot,\cdot)$ e.g. the squared exponenetial covariance function defined in \eqref{42}, $\mathcal{H}_{n_1+n_2}$ be the linear span of
$\left \{ k(\cdot,\bs{x}_i),i=1,...,n_1+n_2 \right \}$,
i.e.
\[\mathcal{H}_{n_{1}+n_{2}}=\left \{ f(\mathord{\cdot}):f(\bs{x})=\sum_{i=1}^{n_{1}+n_{2}}\alpha_{i}k(\bs{x},\bs{x}_{i}), \alpha_{i} \in \mathcal{R} \right \}.\]
We first assume the  true underlying function $\tau_{0} \in \mathcal{H}_{n_{1}+n_{2}}$ then $\tau_{0}(\mathord{\cdot})$ can be expressed as
\[ \tau_{0}(\cdot)=\sum_{i=1}^{n_{1}+n_{2}}\alpha_{i}k(\mathord{\cdot},\bs{x}_{i})\triangleq \bs{K}_{n_{1}n_{2}}(\cdot)\bs{\alpha}.\]
where $\bs{K}_{n_{1}n_{2}}(\cdot)=(k(\mathord{\cdot},\bs{x}_{1}),...,k(\mathord{\cdot},\bs{x}_{n_{1}+n_{2}}))$ and $\bs{\alpha} =(\alpha_{1},\ldots,\alpha_{n_{1}+n_{2}})^{T}$.
By the properties of RKHS, $\left \| \tau_{0} \right \|^{2}_{\bs{K}_{n_{1}n_{2}}}=\bs\alpha^{T}\bs{K}_{n_{1}n_{2}}\bs \alpha$, and $(\tau_0(\bs{x}_{1}),...,\tau_0(\bs{x}_{n_{1}+n_{2}}))^{T}=\bs{K}_{n_{1}n_{2}}\bs \alpha$
where $\bs{K}_{n_{1}n_{2}}=(k(\bs{x}_{i},\bs{x}_{j}))$ is the covariance matrix over $\bs{x}_{i},\quad i=1,\ldots,n_{1}+n_{2}$.

Let $P$ and $\bar{P}$ be any two measures on $\mathcal{F}$, then it yields by the Fenchel-Legendre duality relationship that, for any function $g(\cdot)$ on $\mathcal{F}$,
\begin{equation}
\E_{\bar{P}}[g(\tau)] \leqslant \log \E_{P}[e^{g(\tau)}] +D[\bar{P},P].
\end{equation}
Now in the above inequality let
\begin{enumerate}
\item $g(\tau)$ be $\log p(z_{1},...,z_{n_{1}+n_{2}}|\tau)$ for any $z_{1},...,z_{n_{1}+n_{2}}$ in $\mathcal{Z}$ and  $\tau \in \mathcal{F}$
\item $P$ be the measure induced by $\textup{MGP}(\bs{0},k(\mathord{\cdot},\mathord{\cdot}))$, hence its finite dimensional distribution at $\tau_1,...,\tau_{n_1+n_2}$ is $\mathcal{N}(\bs{0},\hat{\bs{K}}_{n_{1}n_{2}})$ and
\begin{eqnarray*}
\E_{P}[e^{g(\tau)}]&=& \int p(z_{1},...,z_{n_{1}+n_{2}}\mid \tau) p_{n_{1}+n_{2}}(\tau)\textup{d}\tau\\
&=&p_{mgp}(\bs{z})
\end{eqnarray*}
where $\hat{\bs{K}}_{n_{1}n_{2}}$ is defined in the same way as $\bs{K}_{n_{1}n_{2}}$ but the $\bs \theta$ being replaced by its estimator $\hat{\bs \theta}$.

\item $\bar{P}$ be the posterior distribution of $\tau(\mathord{\cdot})$ on $\mathcal{F}$ which has a prior distribution $\textup{MGP} (0,k(\mathord{\cdot},\mathord{\cdot}))$ and normal likelihood $\prod _{i=1}^{n_{1}+n_{2}} N(\hat{z}_i;\tau(\bs{x_{i}}),\sigma^{2})$, where
\begin{eqnarray}\label{*}
\hat{\bs{z}}&\triangleq&\begin{pmatrix}
\hat{z}_{1}\\
\vdots\\
\hat{z}_{n_{1}+n_{2}}
\end{pmatrix}=(\bs{K}_{n_{1}n_{2}}+\sigma^{2}\bs{I})\bs \alpha
\end{eqnarray}
and $\sigma^{2}$ is a constant to be specified. In other words, we assume a model $z=\tau(\bs{x})+\epsilon$ with $\epsilon \sim \mathcal{N}(0,\sigma^{2})$ and $\tau(\mathord{\cdot}) \sim \textup{MGP}(0,k(\mathord{\cdot},\mathord{\cdot}))$, and $\hat{\bs{z}}$ defined by equation \eqref{*} is a set of observations at $\bs{x}_{1},...,\bs{x}_{n_{1}+n_{2}}$. Thus, $\bar{P}(\tau)=p(\tau\mid\hat{\bs{z}},\bs{X}_{n_{1}n_{2}})$ is a probability measure on $\mathcal F$. Therefore, by bivariate CGP regression, the posterior of $(\tau_{1},...,\tau_{n_{1}+n_{2}}) \triangleq (\tau(\bs{x}_{1}),\ldots,\tau(\bs{x}_{n_{1}+n_{2}}))$ is
\begin{eqnarray}\label{c1}
\bar{p}(\tau_{1},...,\tau_{n_{1}+n_{2}})  &\triangleq&  p(\tau_{1},...,\tau_{n_{1}+n_{2}}\mid\hat{\bs{z}},\bs{X}_{n_{1}n_{2}})\nonumber \\
&=&\mathcal{N}(\bs{K}_{n_{1}n_{2}}(\bs{K}_{n_{1}n_{2}}+\sigma^{2}\bs{I})^{-1}\hat{\bs{z}}, \bs{K}_{n_{1}n_{2}}(\bs{K}_{n_{1}n_{2}}+\sigma^{2}\bs{I})^{-1}\sigma^{2})\nonumber \\
&=&\mathcal{N}(\bs{K}_{n_{1}n_{2}}\bs \alpha,\bs{K}_{n_{1}n_{2}}(\mathcal{K}_{n_{1}n_{2}}+\sigma^{2}\bs{I})^{-1}\sigma^{2})) \nonumber\\
&=&\mathcal{N}(\bs{K}_{n_{1}n_{2}}\bs \alpha,\bs{K}_{n_{1}n_{2}}\bs{B}^{-1})
\end{eqnarray}
where $\bs{B}=\bs{I}+\sigma^{-2}\bs{K}_{n_{1}n_{2}}$.
\end{enumerate}

It follows that
\begin{eqnarray}
D[\bar{P},P]&=&\int_{\mathcal F} \log \frac{d\bar{P}}{dP}d\bar{P} \nonumber \\
&=&\int_{\mathcal R^{n_{1}+n_{2}}} \bar{p}(\tau_{1},...,\tau_{n_{1}+n_{2}}) \log \frac {\bar{p}(\tau_{1},...,\tau_{n_{1}+n_{2}})}{\tilde{p}(\tau_{1},...,\tau_{n_{1}+n_{2}})}d \tau_1\cdots d\tau_{n_1+n_2} \nonumber \\
&=&\frac{1}{2}[\log |\widehat{\bs{K}}_{n_{1}n_{2}}|-\log |\bs{K}_{n_{1}+n_{2}}|+\log|\bs{B}|+tr (\widehat{\bs{K}}_{n_{1}+n_{2}}^{-1} \bs{K}_{n_{1}n_{2}}\bs{B}^{-1})+(\bs{K}_{n_{1}n_{2}}\bs \alpha)^{T} \nonumber \\
&&\widehat{\bs{K}}_{n_{1}n_{2}}^{-1}(\bs{K}_{n_{1}n_{2}}\bs \alpha)-(n_1+n_2)] \nonumber \\
&=& \frac{1}{2}[-\log |\widehat{\bs{K}}_{n_{1}n_{2}}^{-1}\bs{K}_{n_{1}+n_{2}}|+\log|\bs{B}|+tr (\widehat{\bs{K}}_{n_{1}+n_{2}}^{-1}\bs{K}_{n_{1}n_{2}}\bs{B}^{-1} )+\left \| \tau_{0} \right \|^{2}_{\bs{K}_{n_{1}n_{2}}} \nonumber \\
&&+\bs \alpha^{T}\bs{K}_{n_{1}n_{2}}(\widehat{\bs{K}}_{n_{1}n_{2}}^{-1}\bs{K}_{n_{1}n_{2}}-\bs{I})\bs \alpha -(n_1+n_2)]. \nonumber
\end{eqnarray}

On the other hand,
\[\E_{\bar{P}}[g(\tau)]=\E_{\bar{P}}[\log p(z_{1},...,z_{n_{1}+n_{2}}|\tau)]=\sum_{i=1}^{n_{1}+n_{2}}\E_{\bar{P}}[\log p(z_{i}|\tau(\bs{x}_{i}))].\]
By Taylor's expansion, expanding $\log p(z_{i}|\tau(\bs{x}_{i}))$ to the second order $\tau_{0}(\bs{x}_{i})$ yields
\begin{eqnarray*}
\log p(z_{i}|\tau(\bs{x}_{i}))&=&\log p(z_{i}|\tau_{0}(\bs{x}_{i}))+\frac{\textup{d}[\log p(z_{i}|\tau(\bs{x}_{i})) ]}{\textup{d}\tau(\bs{x}_{i})}\Bigr|_{\substack{\tau(\bs{x}_{i})=\tau_{0}(\bs{x}_{i})}}(\tau(\bs{x}_{i})-\tau_{0}(\bs{x}_{i})) \\
&& +\frac{1}{2}\frac{\textup{d}^{2}[\log p(z_{i}|\tau(\bs{x}_{i})) ]}{[\textup{d}\tau(\bs{x}_{i})]^{2}}\Bigr|_{\substack{\tau(\bs{x}_{i})=\tilde{\tau}(\bs{x}_{i})}}(\tau(\bs{x}_{i})-\tau_{0}(\bs{x}_{i}))^{2},
\end{eqnarray*}
where $\tilde{\tau}(\bs{x}_{i})=\tau_{0}(\bs{x}_{i})+\lambda(\tau(\bs{x}_{i})-\tau_{0}(\bs{x}_{i}))$ for some $0\leqslant \lambda \leqslant 1$.

For the canonical link function with Convolved GPR , we have
\begin{equation}\label{ic}
\log p(z_{i}|\tau(\bs{x}_{i}))= z_i \log(\bs{U}_{i}^{T}\bs{\beta}+\tau(\bs{x}_i))- (\bs{U}_{i}^{T}\bs{\beta}+\tau(\bs{x}_i)) - \log (z_i!).
\end{equation}
It follows that
\begin{eqnarray*}
\E_{\bar{P}}[\log p(z_{i}|\tau(\bs{x}_{i}))]&=&\log p(z_{i}|\tau_{0}(\bs{x}_{i}))+
(z_i-\exp (\bs{U}_{i}^{T}\bs{\beta}+\tau_0(\bs{x}_i)))
\E_{\bar{P}}[(\tau(\bs{x}_{i})-\tau_{0}(\bs{x}_{i}))] \\
&& -\frac{1}{2}\E_{\bar{P}}[\exp(\bs{U}_{i}^{T}\bs{\beta}+\tilde{\tau}(\bs{x}_i))(\tau(\bs{x}_{i})-\tau_{0}(\bs{x}_{i}))^{2}].
\end{eqnarray*}
Since $\bar{P}(\cdot)$ is the posterior of $\tau(\cdot)$ which has prior $\textup{MGP}(\bs{0},k(\mathord{\cdot},\mathord{\cdot}))$ and normal likelihood $\prod_{i=1}^{n_{1}+n_{2}}\mathcal{N}(\hat{z}_i;\tau(\bs{x}_{i}),\sigma^{2})$, where $\tau(\bs{x}_{i})$ is normally distributed under $\bar{P}$ and it follows from \eqref{c1} that
\begin{eqnarray*}
\tau(\bs{x}_{i}) &\sim& \mathcal {N}(\bs{K}_{n_{1}n_{2}}^{(i)},(\bs{K}_{n_{1}n_{2}}\bs{B}^{-1})_{ii})\\
&=&\mathcal {N}(\tau_{0}(\bs{x}_{i}),(\bs{K}_{n_{1}n_{2}}\bs{B}^{-1})_{ii}))\triangleq \mathcal{N}(\tau_{0i},k_{ii})
\end{eqnarray*}
where $\bs K_{n_{1}n_{2}}^{(i)}$ denotes the $i$th the row of $\bs{K}_{n_{1}n_{2}}$ and $(\bs{K}_{n_{1}n_{2}}\bs{B}^{-1})_{ii}$ is the $i$th diagonal element of $\bs{K}_{n_{1}n_{2}}\bs{B}^{-1}$. Therefore, $\E_{\bar{P}}[\tau(\bs{x}_{i})-\tau_{0}(\bs{x}_{i})]=0$ and
\begin{eqnarray*}
& & \E_{\bar{P}}[\exp (\bs{U}_i^T\bs{\beta}+\tilde{\tau}(\bs{x}_i))(\tau(\bs{x}_{i})-\tau_{0}(\bs{x}_{i}))^{2}] \\
&=& \exp(\bs{U}_i^T\bs{\beta}+\tau_0(\bs{x}_i))\E_{\bar{P}}[e^{ \lambda(\tau(\bs{x}_{i})-\tau_0(\bs{x}_{i}))}(\tau(\bs{x}_{i})-\tau_{0}(\bs{x}_{i}))^{2}]\\
&=& \exp (\bs{U}_i^T\bs{\beta}+\tau_0(\bs{x}_i)+\frac{1}{2}\lambda^{2}k_{ii})(\lambda^{2}k_{ii}+1)k_{ii} \leqslant \tilde{\delta}k_{ii}
\end{eqnarray*}
since the covariance function is bounded. Here $\tilde{\delta}$ is a generic positive constant.
Thus, we have
\[-\sum_{i=1}^{n_{1}+n_{2}}\E_{\bar{P}}[\log p(z_{i}|\tau(\bs{x}_{i}))] \leqslant -\sum_{i=1}^{n_{1}+n_{2}}\log p(z_{i}|\tau_{0}(\bs{x}_{i}))+\frac{\tilde{\delta}}{2}  tr(\bs{K}_{n_{1}n_{2}}\bs{B}^{-1}).\]
$i.e.$
\[\log p_{0}(z_{1},...,z_{n_{1}+n_{2}})\leqslant \E_{\bar{P}}[g(\tau)]+\frac{\tilde{\delta}}{2}  tr(\bs{K}_{n_{1}n_{2}}\bs{B}^{-1}).\]
Combining the bounds gives
\begin{eqnarray}\label{1a}
   && -\log p_{mgp}(z_{1},...,z_{n_{1}+n_{2}})+\log p_{0}(z_{1},...,z_{n_{1}+n_{2}})\nonumber \\
  \leqslant&&-\log \E_{P}[e^{g(\tau)}]+\E_{\bar{P}}[g(\tau)]+\frac{\tilde{\delta}}{2}  tr(\bs{K}_{n_{1}n_{2}}\bs{B}^{-1}) \nonumber\\
\leqslant &&D[\bar{P},P]+\frac{\tilde{\delta}}{2}  tr(\bs{K}_{n_{1}n_{2}}\bs{B}^{-1})\nonumber \\
=&&\frac{1}{2}[-\log |\widehat{\bs{K}}_{n_{1}n_{2}}^{-1}\bs{K}_{n_{1}n_{2}}|+\log|\bs{B}|+tr (\widehat{\bs{K}}_{n_{1}n_{2}}^{-1}\bs{K}_{n_{1}n_{2}}\bs{B}^{-1} +\tilde{\delta}\bs{K}_{n_{1}n_{2}}\bs{B}^{-1})+\left \| \tau_{0} \right \|^{2}_{\bs K_{n_{1}n_{2}}} \nonumber \\
&&+\bs \alpha^{T}\bs{K}_{n_{1}n_{2}}(\widehat{\bs{K}}_{n_{1}n_{2}}^{-1}\bs{K}_{n_{1}n_{2}}-\bs{I})\bs \alpha -(n_1+n_2)].
\end{eqnarray}

Since the covariance function is continuous in $\bs \theta$ and $\hat{\bs \theta}_{n_{1}+n_{2}}\rightarrow  \bs \theta$  and we have $\widehat{\bs{K}}_{n_{1}n_{2}}\bs{K}_{n_{1}n_{2}}-\bs{I} \rightarrow 0$ as $n_{1} \rightarrow \infty$ and $n_{2} \rightarrow \infty$, hence $n_{1}+n_{2} \rightarrow \infty$. Therefore there exist some positive constants $C$ and $\epsilon$ such that
\[-\log |\widehat{\bs{K}}_{n_{1}n_{2}}^{-1}\bs{K}_{n_{1}n_{2}}|<C, \quad \bs \alpha^{T}\bs{K}_{n_{1}n_{2}}(\widehat{\bs{K}}_{n_{1}n_{2}}^{-1}\bs{K}_{n_{1}n_{2}}-\bs{I})\bs \alpha < C, \]
\[tr(\widehat{\bs{K}}_{n_{1}+n_{2}}^{-1}\bs{K}_{n_{1}n_{2}}\bs{B}^{-1}) < tr((\bs{I}+\epsilon\bs{K}_{n_{1}n_{2}})\bs{B}^{-1}),\]
since the covariance function is bounded.

Thus the right hand side (RHS)  of \eqref{1a}\[ < \frac{1}{2}\left \| \tau_{0} \right \|^{2}_{\bs{K}_{n_{1}n_{2}}} +\frac{1}{2}  [2C+\log |\bs{B}|+tr((\bs{I}+(\epsilon+\tilde{\delta})\bs{K}_{n_{1}n_{2}})\bs{B}^{-1})-(n_1+n_2)].\]

Note that thee above inequality holds for all $\sigma^{2} > 0$, thus letting $\sigma^{2}=(\epsilon +\tilde{\delta})^{-1}$ and $\delta=\epsilon+\tilde{\delta}$ yields that the RHS of \eqref{1a} becomes
\[\frac{1}{2}\left \| \tau_{0} \right \|^{2}_{\bs{K}_{n_{1}n_{2}}} +\frac{1}{2}  \log(\bs{I}+\delta\bs{K}_{n_{1}n_{2}})+C.\]
Thus we have
\begin{eqnarray}\label{2}
-\log p_{mgp}(z_{1},...,z_{n_{1}},z_{n_{1}+1},...,z_{n_{1}+n_{2}}) &\leqslant& -\log p_{0}(z_{1},...,z_{n_{1}},z_{n_{1}+1},...,z_{n_{1}+n_{2}})+\frac{1}{2}\left \| \tau_{0} \right \|^{2}_{\bs{K}_{n_{1}n_{2}}} + \nonumber\\
 &&\frac{1}{2}  \log(\bs{I}+\delta\bs{K}_{n_{1}n_{2}})+C
\end{eqnarray}
for any $\tau_{0}(\mathord{\cdot}) \in \mathcal{H}_{n_{1}+n_{2}}$.

Taking infimum on RHS of \eqref{2} over $\tau_{0}$ and applying \textit{Representer Theorem}, we obtain
\begin{eqnarray*}
-\log p_{mgp}(z_{1},...,z_{n_{1}+n_{2}}) +\log p_{0}(z_{1},...,z_{n_{1}+n_{2}})\\
\leqslant \frac{1}{2}\left \| \tau_{0} \right \|^{2}_{\bs{K}_{n_{1}n_{2}}} +\frac{1}{2}  \log(\bs{I}+\delta\bs{K}_{n_{1}n_{2}})+C
\end{eqnarray*}
for all $\tau_{0}(\mathord{\cdot}) \in \mathcal{H}_{n_{1}+n_{2}}$. The proof is complete.
\end{proof}
\begin{proof}[Proof of Theorem 1]
It follows from the definition of information consistency that
\begin{displaymath}
  D[p_{0}(\bs{z}),p_{mgp}(\bs{z})] = \int p_{0}(z_{1},\ldots,z_{n_{1}+n_{2}})\log\frac {p_{0}(z_{1},\ldots,z_{n_{1}+n_{2}})}{p_{mgp}(z_{1},\ldots,z_{n_{1}+n_{2}})}
  d z_{1}\cdots d z_{n_{1}+n_{2}}.
\end{displaymath}
Applying Lemma 1 we obtain that
\begin{eqnarray}\label{con4}
\frac{1}{n_{1}+n_{2}}\E_{\bs{X}_{n_{1}n_{2}}}\left ( D[p_{0}(\bs{z}),p_{mgp}(\bs{z})] \right )&\leqslant& \frac{1}{2(n_{1}+n_{2})}\left \| \tau_{0} \right \|^{2}_{\bs{K}_{n_{1}n_{2}}} +\frac{1}{2(n_{1}+n_{2})} \E_{\bs{X}_{n_{1}n_{2}}} \log(\bs{I}\nonumber\\
&&+\delta\bs{K}_{n_{1}n_{2}})+\frac{C}{n_{1}+n_{2}},
\end{eqnarray}
where $\delta$ and $C$ are some positive constants. Theorem 1 follows from \eqref{con4}.
\end{proof}

\textbf{Remark 2}\label{1}
Lemma 1 requires that the estimator of the coefficients $\bs \beta$ and hyper-parameters $\bs \theta$ are consistent. Yi et al. (2011) provided that the empirical Bayesian estimator of hyper-parameters $\bs \theta$ as $ n \rightarrow \infty $ under certain regularity. The estimator $\bs \beta$ and $\bs \theta$ for bivariate Poisson regression with CGP priors are consistent under certain regularity, if $n=n_{1}+n_{2}$, where $n_{1} \rightarrow \infty \quad  and \quad n_{2} \rightarrow \infty$.

\textbf{Remark 3}
 Some specific results of the regret term $R=\E_{\bs{X}_{n_{1}n_{1}}}(\log |\bs{I}+\delta\bs{K}_{n_{1}n_{2}} |)$ as follows :
\begin{enumerate}
\item if $k(\bs{x}_{i},\bs{x}_{j})=\bs{x}_{i}^{T}\bs{x}_{j}$, i.e. a linear covariance kernel, and the covariate distribution $\bs{u}(\bs{x})$ has bounded support, then
\begin{eqnarray*}
\E_{\bs{X}_{n_{1}n_{1}}}(\log |\bs{I}+\delta\bs{K}_{n_{1}n_{2}} |)=O(\log(n_{1}+n_{2}));
\end{eqnarray*}
\item if $\bs{u}(\bs{x})$ is normal and the covariance functions are  the squared exponential form, then
\begin{eqnarray*}
\E_{\bs{X}_{n_{1}n_{1}}}(\log |\bs{I}+\delta\bs{K}_{n_{1}n_{2}} |)=O((\log(n_{1}+n_{2}))^{p+1});
\end{eqnarray*}
\item if $\bs{u}(\bs{x})$ is bounded support and the covariance functions are Matern, then
\begin{eqnarray*}
\E_{\bs{X}_{n_{1}n_{1}}}(\log |\bs{I}+\delta\bs{K}_{n_{1}n_{2}} |)=O((n_{1}+n_{2})^{p/(2v+p)}(\log (n_{1}+n_{2})^{2v/(2v+p)}));
\end{eqnarray*}
\item if  covariance functions are mixed between squared exponential and Matern,  then
\begin{eqnarray*}
\E_{\bs{X}_{n_{1}n_{1}}}(\log |\bs{I}+\delta\bs{K}_{n_{1}n_{2}} |)=O((n_{1}+n_{2})^{p/(2v+p)}(\log (n_{1}+n_{2})^{2v/(2v+p)})).
\end{eqnarray*}
\end{enumerate}
Thus  the information consistency in the proposed model is achieved for all of the above cases.

\end{document}